\documentstyle[aps]{revtex}
\font\eightrm=cmr8
\def\bxk{{\scriptstyle({\eightrm x} \scriptstyle)}}
\def\bxpk{{\scriptstyle({\eightrm x}^{\eightrm\prime} \scriptstyle)}}

\def\lxr{{\scriptstyle[{\eightrm x} \scriptstyle]}}

\def\bzk{{\scriptstyle({\eightrm z} \scriptstyle)}}
\def\bt0zk{{\scriptstyle({\eightrm t}_{\eightrm 0}, {\eightrm{\bf z}})}}
\def\mmbox#1#2{\vcenter{\hrule \hbox{\vrule height#2in
                \kern#1in \vrule} \hrule}}

\input{epsf}
\begin{document}
\draft
\begin{flushright} TOKAI-HEP/TH-0006 \end{flushright}
\vspace{0.5cm}
\centerline{\bf Relativistically covariant formulation}
\centerline{\bf of the canonical theory of classical fields I}
\vspace{1.0cm}
\centerline{Hiroshi Ozaki\footnote{Email address: ozaki@keyaki.cc.u-tokai.ac.jp}}
\vspace{0.5cm}
\centerline{\small {\it Department of Physics, Tokai University, 1117 Kitakaname, Hiratsuka 259-1292, Japan}}
\vspace{1.0cm}
\begin{abstract}
An explicit Lorentz covariant formulation of the canonical theory for classical fields is established on a
space-like hypersurface. Hamilton's equations and
a Poisson bracket are defined on the space-like hypersurface. The Poisson bracket
relations between total momentum and total angular momentum satisfies the  
Poincar{\'e} algebra. It is shown that our Poisson bracket
has the same symplectic structure that was brought in the covariant symplectic approach.  
\end{abstract}
\pacs{ }

\section{Introduction}

This paper presents a canonical formalism for classical fields to be coherent with explicit Lorentz covariance.
The usual canonical formalism of the classical field theory was developed on the basis of non-relativistic concepts 
which make a sharp distinction between space and time. 
The field equations are written in the Hamiltonian form, which show how the dynamical systems of fields 
change with time. However, the disadvantage of such a formalism is that its Lorentz covariance is not evident, 
since the Hamiltonian form of the equation is linked to a definite choice of Lorentz frame of reference. 
The definition of momentum canonically conjugate to a field also depends on the frame of reference. 
The canonical Poisson bracket is also defined to involve two field variables at different points in space 
but at the same time. Thus the transformation property under any proper 
homogeneous Lorentz transformation are not transparent.
This defect transfers directly into the canonial formalism in quantum field theory. 
Since equal-time commutation relations between canonical field operators and momenta 
are not derived within the framework of quantum field theory, 
the equal-time canonical Poisson bracket prescribes the equal-time commutation laws. 
Fortunately, in quantum field theory, the field equation or the spectral representation 
brings the equal-time commutation relations into the four-dimensional commutators, 
which show the explicit Lorentz covariance. It may make us to believe that this can be done 
because the whole theory is constructed in the completely covariant manner, 
and it is not essential whether the explicit Lorentz covariance maintains from the outset. 
However, the procedure of the equal-time quantization forming a corner-stone of quantum field theory 
does not appear reasonable due to lack of the relativistic covariance.

 One way out of the difficulty is to refine the canonical formalism for  classical fields 
 to be coherent with Lorentz covariance.  Here, I will take a method of using the space-like hypersurface. 
 The refinement will make us free from  restriction of equal-time, and make the procedure of the canonical quantization 
 to be covariant throughout.


I use natural units throughout with $\hbar = c = 1$.
My conventions for relativity follow Bjorken and Drell~\cite{BD}(1964). 
I use a metric tensor of space-time  $g_{\mu \nu} = {\rm diag.}(+---)$ with Greek indices 
ranging over $0,1,2,3$ or $t,x,y,z$. Latin indices --i,j,k etc.--range over $1,2,3$ 
and represent coordinates etc. in three-dimensional space. 
The space-time coordinates are denoted by the contravariant four-vector:
$$x^{\mu} = (x^{0}, x^{1}, x^{2}, x^{3})=( t, x, y, z).$$
The covariant four-vector $x_{\mu}$ is obtained by:
$$x_{\mu}=(x_{0}, x_{1}, x_{2}, x_{3})=( t, -x, -y, -z)= g_{\mu \nu}  x^{\nu}.$$

\section {The canonical formalism on a space-like hypersurface \protect\\ for classical fields}

The usual canonical formalism for classical fields contains four non-covariant quantities:
(1) the momentum canonically conjugate to a field
(2) the Hamiltonian density
(3) Hamilton's equations
(4) the canonical Poisson bracket.

The momentum canonically conjugate to a field and the Hamiltonian density have time derivative in definition. 
The canonical Poisson bracket is defined to involve two field variables at the same time. 
Since time is sharply distinct with space in the usual canonical formalism, 
(1)$\sim$(4) do not maintain form invariance under any proper homogeneous Lorentz transformation. 
We wish to define them in the Lorentz covariant manner so that those have the same form 
in all Lorentz frames of reference.  Here, we will take a method of using the space-like hypersurface.
%
\subsection{The space-like hypersurface}

The ordinary classical field theory formulates a physical law by describing 
how the dynamical systems of classical fields $\phi _A  (A=1,2, \cdots ,N)$  changes with time. 
The spatial distribution of every field at time $t_0$ is connected with a hyper-plane
associated with some frame of reference. 
It will change into another at any time $t$ if every field was disturbed by an interaction. 
Then a new distribution will be connected with another hyper-plane, located over the hyper-plane at $t_0$. 
We can evaluate the change of those fields when the limit
$$
{{\partial \phi _A\bxk} \over {\partial t}}=\mathop {\rm {lim}}\limits_{t\to t_0}
{{\phi _A{\scriptstyle(t,{\bf x}\scriptstyle)}-\phi _A{\scriptstyle(t_0,{\bf x}\scriptstyle)}} \over {t-t_0}},
$$
exists. 
We may regard that this derivative is defined between a pair of hyper-planes in space-time. 
The hyper-planes belong to the so-called space-like hypersurfaces. 
A set of points is the space-like hypersurface on which every pair of points on the surface 
is separated by a space-like interval to insure the micro-causality.

\begin{figure}
\begin{center}
\epsfxsize=4cm
\epsfbox{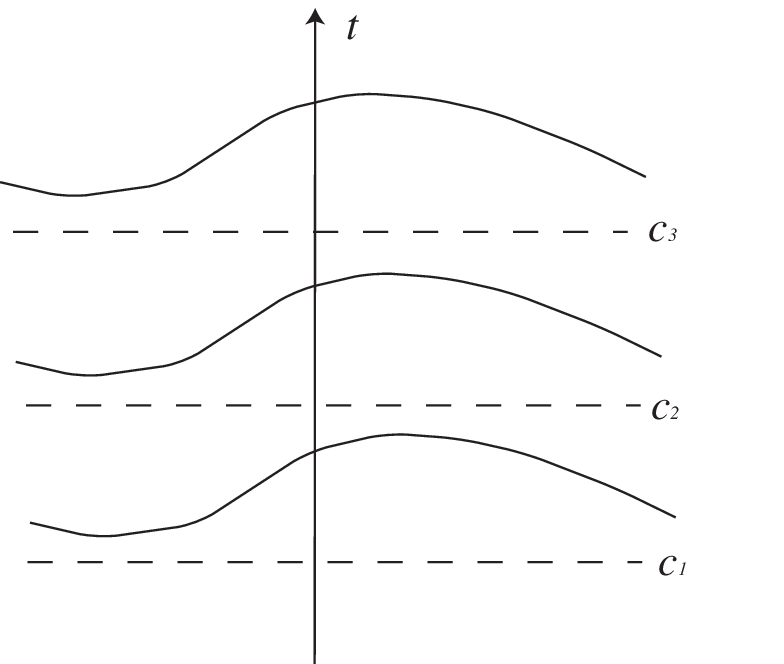}
\end{center}
\caption{The slicing of space-time into a one-parameter family 
of space-like hypersurfaces ${c_1, c_2, c_3}$.}
\label{fig1}
\end{figure}

Let $t$ be a space-like hypersurface of the form
$$
t=\sigma {{\scriptstyle({\eightrm x,y,z} \scriptstyle)}}=g{{\scriptstyle({\eightrm x,y,z} \scriptstyle)}}+c,
$$
where $c$ is a real parameter, and $g$ is a function of space-coordinates.
For a fixed function $g$,  successive values of $c$ makes a foliation of space-time 
into space-like sections, shown in Fig.\ \ref{fig1}, each being a complete (global) Cauchy hypersurface.  
For a fixed parameter $c$, continuous change of $g$ makes a deformation of the space-like hypersurface
as shown in Fig.\ \ref{fig2}.
The function $\sigma$ will alter under any proper homogeneous Lorentz transformation, 
but the transformed function also remains as a space-like hypersurface.  
Thus the property of the space-like hypersurface is compatible 
with the requirement of the Lorentz covariance.

If we choose especially $g$ to be zero, the space-like hypersurface coincides with usual time as stated above. 
Then the space-like hypersurface is restricted to be a plane surface, 
which means that every pair of points on the plane is in equal time. 

We follow Schwinger~\cite{Schwinger} in defining the three-dimensional volume element at a point $x$ 
on a space-like hypersurface by $d\Sigma^{\mu} \bxk = n^{\mu}\bxk d\Sigma$ , 
where $n^{\mu}\bxk$ is a time-like unit vector $(n^\mu \bxk n_\mu \bxk = 1)$ and $d\Sigma$ 
is the numerical measure of the surface element.
The coordinate derivatives can be decomposed into components normal$(n)$ 
and tangential$(t)$ to a space-like hypersurface:
$$\partial_{\mu} = n_{\mu}\bxk \partial_{n} + {\partial_{t}}_{\mu},$$
$$\partial_{n}  = n_{\mu}\bxk \partial^{\mu},$$
$${\partial_{t}}_{\mu} = P_{\mu\nu} \partial^{\nu},$$
where $P_{\mu\nu}=g_{\mu \nu} - n_{\mu}\bxk n_{\nu}\bxk$ is the  projection tensor into the space-like hypersurface.
If we take a flat space-like hypersurface, we get
$$\partial _n={\partial  \over {\partial t}},\ \ \ \partial _{t\mu }=(0, -\nabla) $$
because of $n^{\mu}=(1,0,0,0)$ for the flat surface.

\subsection {The Lorentz covariant momenta canonically conjugate to classical fields}

Let us consider a set of classical fields ${\phi_{A}}\bxk$ at a space-time point $x$ 
with the Lagrangian density ${\cal L} (\phi_{A}\bxk, \partial_{\mu}{\phi_{A}}\bxk) = {\cal L} \lxr$. 
The Lagrangian density is well-defined not only in all space-time points but also on any space-like hypersurface. 
Then we introduce a set of Lorentz covariant momenta canonically conjugate to $\phi_{A}\bxk$ on a space-like hypersurface:
\begin{equation}
{\Pi_{A}} \bxk = n_{\mu} \bxk{\partial^{R} {\cal L}\lxr \over \partial [\partial_{\mu} {\phi_{A}}\bxk]}. \label{eq:mom1}
\end{equation}
The momenta (\ref{eq:mom1}) agree with the usual non-covariant momenta when we take a flat space-like hypersurface:
$$
\pi _A \bxk={{\partial ^R{\cal L}\lxr} \over {\partial \dot \phi _A \bxk}}.
$$
Thus the definition of the Lorentz covariant momenta is a natural extension 
of the conventional momenta defined at a particular time. 

We must often deal with constrained gauge systems 
in which a momentum canonically conjugate to a classical field vanishes. 
We know a practical case of this kind is provided
by the electromagnetic field, in fact, the momentum to the scalar potential at any point
vanishes. This attributes to the invariance of the original Lagrangian density under the gauge
transformation. The constrained Hamiltonian reformulation by Dirac~\cite{Dirac} is widely used to avoid
having irregularities in the theory. Another way of escaping from the irregularities
is to add the gauge fixing term ${\cal L}_{\rm {GF}}\lxr$
to the original gauge invariant Lagrangian density${\cal L}_s \lxr$:
${\cal L}\lxr = {\cal L}_s\lxr + {\cal L}_{\rm {GF}}\lxr$.
The ${\cal L}\lxr$ leads to the field equation
without the (local) gauge invariant term and gives nonvanishing momentum.
Although there remains an uncertainty to the choice of the gauge fixing term, 
this method is simpler than the Dirac's.
We shall take a method of using the Lagrangian density ${\cal L}\lxr$ containing
the covariant gauge fixing term when we must deal with constrained gauge systems.
%
\subsection {The Hamiltonian} 

With the help of the Lorentz covariant momenta, we can apply Legendre's transformation 
at a point $x$ on a space-like hypersurface $\sigma$ to get the new function, 
which is now denoted by ${\cal H}\lxr$:
\begin{equation}
{\cal H }{\lxr }= {\Pi_{A}}\bxk \partial_{n} \phi_{A} \bxk - {\cal L}\lxr. \label{eq:hamden}
\end{equation}
As stated above, we shall avoid having irregularities in the theory by the covariant gauge fixing, 
but it is not yet obvious whether the equations (\ref{eq:mom1}) are solvable for the 
$\partial_n \phi_{A}\bxk$. Here we will be restricted to the solvable systems.
Then we can express the new function ${\cal H}\lxr$ in terms of
the Lorentz covariant momenta $\Pi_A \bxk$ by solving the equation (\ref{eq:mom1}) for
the $\partial_n \phi_{A}\bxk$. 
We thus obtain
$$
{\cal H} \lxr= {\cal H}({\phi_A}\bxk, {\partial_{t\mu}} {\phi_A}\bxk, {\Pi_A}\bxk, {\partial_{t\mu}}{\Pi_A}\bxk).
$$
We will notice that ${\cal H}{\lxr }$ is a Lorentz scalar density introduced by Matthews~\cite{Matthews}:
${\cal H }\lxr  = n^{\mu}\bxk n^{\nu}\bxk T_{\mu \nu}\lxr$. 
Let $H[\sigma]$ be as an integral of ${\cal H}\lxr$ over $\sigma$
\begin{eqnarray}
H [\sigma] &=& \int_{\sigma} d\Sigma {\cal H}{\lxr } \nonumber\\
           &=& \int_{\sigma} d\Sigma^{\mu}\bxk n^{\nu}\bxk {T_{\mu \nu}}\lxr.
               \label{eq:hamiltonian}
\end{eqnarray}
On substituting $\delta x_{\mu} = \epsilon n_{\mu}\bxk$ ($\epsilon$ is an infinitesimal parameter) 
and $\delta \phi_A \bxk = 0$ in (A.3), we obtain $J[\sigma] = -\epsilon H[\sigma]$. 
This shows that $H[\sigma]$ is a conserved charge, so $H[\sigma]$
does not depend on the choice of $\sigma$. Thus we can drop the label $[\sigma]$ and
may abbreviate (\ref{eq:hamiltonian}) as
\begin{equation}
H  = \int_{\sigma} d\Sigma {\cal H}. \label{eq:totalham}
\end{equation}
Further, ${\cal H}\lxr$ reduces to the ordinary Hamiltonian density ${\bar{\cal H}}{\lxr }$:
$${\bar{\cal H}}{\lxr } = {\pi_{A}}\bxk \dot \phi_{A} \bxk - {\cal L}\lxr$$
when $\sigma$ is flat.
Thus we may call the new-function ${\cal H}{\lxr }$ the total Hamiltonian density and 
$H$ the total Hamiltonian.
%
\subsection {Hamilton's equations}

The variation of the total Hamiltonian (\ref{eq:totalham}) is equal to
$$\delta H = \int_\sigma  {d\Sigma \ [\delta \Pi _A \cdot \partial _n\phi _A 
-({{\partial ^R \cal L} \over {\partial \phi _A }}}
-\partial _\mu {{\partial ^R \cal L} \over {\partial [\partial _\mu \phi _A ]}})
\delta \phi _A -\partial _{t\mu}({{\partial ^R \cal L} \over {\partial [\partial _\mu \phi _A ]}}\delta \phi _A )].$$
Since the total Hamiltonian density (\ref{eq:hamden}) is a function of classical fields and Lorentz covariant momenta:
$${\cal H} = {\cal H}({\phi_A}, {\partial_{t\mu}} {\phi_A}, {\Pi_A}, {\partial_{t\mu}}{\Pi_A}),$$
the variation of (\ref{eq:totalham}) can also be written as
\begin{eqnarray*}
\delta H &=& \int_\sigma  d\Sigma \ [({{\partial ^R {\cal H}} \over {\partial \phi _A}}
-\partial _{t\mu} {{\partial ^R {\cal H}} \over {\partial [\partial _{t\mu} \phi _A]}})\delta \phi _A 
+(-)^{\left| A  \right|}\delta \Pi _A ({{\partial ^R {\cal H}} \over {\partial \Pi _A}}
-\partial _{t\mu} {{\partial ^R {\cal H}} \over {\partial [\partial _{t\mu} \Pi _A ]}}) \\
  & &{ }+\partial _{t\mu }{{{\partial ^R {\cal H}} \over {\partial [\partial _{t\mu}\phi _A]}}\delta \phi _A 
  +(-)^{\left| A  \right|}\delta \Pi _A \partial _{t\mu }{{\partial ^R {\cal H}} \over {\partial [\partial _{t\mu}\Pi _A]}}}],
\end{eqnarray*}
where the symbols $\partial^R/ \partial \xi_A$  and $(-)^{\vert A \vert}$  are given in Appendix.

We assume canonical variables decrease fast enough at space-like infinity. 
Each of the boundary terms disappears because of $\delta{\phi_{A}} \rightarrow 0; 
\delta{\Pi_{A}} \rightarrow 0$ for the boundary $\partial \sigma$.
Thus we get Hamilton's equations in the Lorentz covariant form
\begin{eqnarray}
(-)^{\left| A  \right|}\partial _n\phi _A &=&{{\partial ^R \cal H} \over {\partial \Pi _A }}
-\partial _{t\mu}{{\partial ^R \cal H} \over {\partial [\partial _{t\mu}\Pi _A ]}},\label{eq:cano1} \\
-\partial _n\Pi _A &=&{{\partial ^R \cal H} \over {\partial \phi _A }}
-\partial _{t\mu}{{\partial ^R \cal H} \over {\partial [\partial _{t\mu}\phi _A ]}}, \label{eq:cano2}
\end{eqnarray}
with the help of the Euler-Lagrange equation
$$
{\partial^{R} {\cal L}\over \partial{\phi_{A}}}
-{{\partial_{\mu}}}{{\partial^{R} {\cal L}\over\partial[{\partial_{\mu}}{\phi_{A}}]}} = 
{\partial^{R} {\cal L}\over \partial{\phi_{A}}}
-{{\partial_{t\mu}}}{{\partial^{R} {\cal L}\over\partial[{\partial_{t\mu}}{\phi_{A}}]}} - \partial_{n} {\Pi_{A}}
=0.
$$

Here, let us introduce the functional derivatives on the space-like hypersurface.
Take a surface-integral
\begin{eqnarray}
F[\sigma;\phi _A,\Pi _A]&=&\int_{\sigma} {d\Sigma ^\nu \bxk}\ f_\nu 
(\phi _A \bxk,\partial _{\mu t}\phi _A(x),\Pi _A \bxk,\partial _{\mu t}\Pi _A \bxk)\nonumber\\
                        &=&\int_{\sigma} {d\Sigma \ {\cal F}
(\phi _A,\partial _{\mu t}\phi _A,\Pi _A,\partial _{\mu t}}\Pi _A), \label{eq:sfint}
\end{eqnarray}
where
$d\Sigma ^\nu f_{\nu}=d\Sigma (n\!\cdot\! f) \equiv d\Sigma {\cal F}.$
We pass to an infinitesimal displacement of canonical variables by making 
\begin{eqnarray*}
\phi _A \bxk &\to& \phi _A \bxk+\varepsilon\chi _A \bxk, \\
 \Pi _A \bxk &\to& \Pi _A \bxk+\varepsilon\chi _A \bxk ,
\end{eqnarray*}
where each of $\chi_A \bxk$ is a function.
Then the functional derivatives of (\ref{eq:sfint}) on the space-like hypersurface are defined as
\begin{eqnarray}
\int_\sigma  {d\Sigma {{\tilde \delta F[\sigma ]} \over {\tilde \delta \phi _A\bzk}}}
\ \chi_A  \bzk &\equiv& \mathop {{\rm lim}}\limits_{\varepsilon \to 0}
{{F[\sigma ;\phi _A +\varepsilon \chi_A  ,\Pi _A ]-F[\sigma ;\phi _A,\Pi _A ]} \over \varepsilon },\label{eq:fder1}\\
  \int_\sigma  {d\Sigma {{\tilde \delta F[\sigma ]} \over {\tilde \delta \Pi _A\bzk}}}
\ \chi_A  \bzk &\equiv& \mathop {{\rm lim}}\limits_{\varepsilon \to 0}
{{F[\sigma ;\phi _A ,\Pi _A +\varepsilon \chi_A  ]-F[\sigma ;\phi _A,\Pi _A ]} \over \varepsilon },\label{eq:fder2}
\end{eqnarray}
where $F[\sigma ]$ is short for $F[\sigma ;\phi _A,\Pi _A]$. 
Note that we do not take the summation on $A$ in the left-hand side of (\ref{eq:fder1}) and (\ref{eq:fder2}).
To get a practical form for functional derivatives,
we will use a $\delta$-function $\delta _\sigma (x-z)$~\cite{SW2} on the space-like hypersurface:
$$\int_\sigma  {d\Sigma \ }{\delta _\sigma}(x-z)\xi _A\bzk=\xi _A\bxk,$$
where $x$ is a point on $\sigma$.
It follows that
$$
{{\tilde \delta F[\sigma ]} \over {\tilde \delta \phi _A \bzk}}
={{\partial ^R{\cal F}} \over {\partial \phi _A \bzk}}-\partial _{t\mu} 
{{\partial ^R{\cal F}} \over {\partial [\partial _{t\mu} \phi _A \bzk]}},
$$
$$
{{\tilde \delta F[\sigma ]} \over {\tilde \delta \Pi _A \bzk}}
={{\partial ^R{\cal F}} \over {\partial \Pi _A \bzk}}-\partial _{t\mu} 
{{\partial ^R{\cal F}} \over {\partial [\partial _{t\mu} \Pi _A \bzk]}}.$$
With the help of the functional derivatives, we may express Hamilton's equations (\ref{eq:cano1}) and (\ref{eq:cano2})
more compact form
\begin{eqnarray}
\left( - \right)^{\left| A \right|}\partial _n\phi _A \bzk
&=&{{\tilde \delta H} \over {\tilde \delta \Pi _A \bzk}},\label{eq:cano1a} \\
-\partial _n\Pi _A \bzk
&=&{{\tilde \delta H} \over {\tilde \delta \phi _A \bzk}}. \label{eq:cano2a}
\end{eqnarray}

%
\subsection {The four-dimensional Poisson bracket}

Now, we define a Poisson bracket on the space-like hypersurface. 
Any two Lorentz covariants $F[\sigma]$ and $G[\sigma]$ 
being the function of canonical field variables and momenta on $\sigma$ have a Poisson bracket 
which we shall denote by $\left[ {F[\sigma],G[\sigma]} \right]_c$, defined by
\begin{equation}
\left[ {F[\sigma],G[\sigma]} \right]_c\!\!
=\int_\sigma  \!\!\!{d\Sigma}\!\left( {{{\tilde \delta F[\sigma]} \over {\tilde \delta \phi _A \bzk}}
{{\tilde \delta G[\sigma]} \over {\tilde \delta \Pi _A \bzk}}
-(-)^{\left| A \right|}{{\tilde \delta F[\sigma]} \over {\tilde \delta \Pi _A \bzk}}
{{\tilde \delta G[\sigma]} \over {\tilde \delta \phi _A \bzk}}} \right). \label{eq:pb}
\end{equation}
The subscript ``$c$" means that the bracket is defined classically. 
The algebraic properties are the same as those of generalized Poisson bracket~\cite{HT},
for example,
\begin{eqnarray}
\left[ {A,B} \right]_c &=& -(-)^{\left| A \right|\left| B \right|}[B,A]_c,\label{eq:pbalg1} \\
\left[ {A+B,C} \right]_c &=& \left[ {A,C} \right]_c+\left[ {B,C} \right]_c,\label{eq:pbalg2} \\
\left[ {A,BC} \right]_c &=& \left[ {A,B} \right]_c C+(-)^{\left| A \right|\left| B \right|}B\left[ {A,C} \right]_c.
\label{eq:pbalg3}
\end{eqnarray}
A number or physical constant may be counted as a special case of canonical variables, 
and has the property that its Poisson bracket with anything vanishes.
The Poisson bracket on $\sigma$ also satisfies the generalized Jacobi identity
\begin{eqnarray}
\left[ {\left[ {A,B} \right]_c,C} \right]_c 
  &+&(-)^{\left| C \right|(\left| A \right|+\left| B \right|)}\left[ {\left[ {C,A} \right]_c,B} \right]_c \nonumber\\
  &+&(-)^{\left| A \right|(\left| B \right|+\left| C \right|)}
\left[ {\left[ {B,C} \right]_c,A} \right]_c=0.\label{eq:pbalg4}
\end{eqnarray}
It is obvious from its definition that our Poisson bracket is covariant under 
any proper homogeneous Lorentz transformation. 
When the space-like hypersurface $\sigma$ is flat, $\sigma = t_0$,  and all canonical variables 
are commuting c-numbers, our Poisson bracket reduces to
$$
\left[ {F[t _0],G[t _0]} \right]_c=\int \!{d^3 z} \left( {{{ \delta F[t_0]} \over {\delta \phi _A\bt0zk}}
{{\delta G[t _0]} \over {\delta \pi _A
\bt0zk}}-{{\delta F[t _0]} \over {\delta \pi _A
\bt0zk}}{{\delta G[t _0]} \over {\delta \phi _A\bt0zk}}} \right).
$$
This is just the ordinary Poisson bracket. 
Thus, the Poisson bracket on $\sigma$ is also a natural extension of the ordinary Poisson bracket. 
Hereafter, we call our Poisson bracket the four-dimensional Poisson bracket.

Let us define an infinitesimal canonical transformation between field variables 
and momenta on a space-like hypersurface as
\begin{eqnarray}
& &\bar \delta \phi _A\bxk=\phi ^{\prime}_A\bxk-\phi_A \bxk
=\varepsilon (-)^{\left| A \right|(\left| \Gamma  \right|+1)}
{{\tilde \delta \Gamma [\sigma ]} \over {\tilde \delta \Pi _A\bxk}},\label{eq:ctransf1} \\
& &\bar \delta \Pi _A\bxk=\Pi ^{\prime}_A\bxk-\Pi_A \bxk
=-\varepsilon (-)^{\left| A \right|\left| \Gamma  \right|}
{{\tilde \delta \Gamma [\sigma ]} \over {\tilde \delta \phi _A\bxk}},\label{eq:ctransf2}
\end{eqnarray}
where the additional
lavel $\prime$ is to distinguish canonical variables from previous ones. 
$\Gamma [\sigma]$ is the generating function of the transformation given by
$$\Gamma [\sigma ]=
\int_\sigma {d\Sigma \ {\cal G}}.
$$
The four-dimensional Poisson bracket is invariant under the infinitesimal canonical transformation 
\begin{eqnarray*}
\left[ {F[\sigma],G[\sigma]} \right]_c &=&\int_{\sigma} \!\!\!{d\Sigma }
\left( {{{\tilde \delta F[\sigma]} \over {\tilde \delta \phi _A\bzk}}
{{\tilde \delta G[\sigma]} \over {\tilde \delta \Pi _A\bzk}}-(-)^{\left| A \right|}
{{\tilde \delta F[\sigma]} \over {\tilde \delta \Pi _A\bzk}}
{{\tilde \delta G[\sigma]} \over {\tilde \delta \phi _A\bzk}}} \right) \\
&=&\int_{\sigma} \!\!\!{d\Sigma }\left( {{{\tilde \delta F[\sigma]} \over {\tilde \delta \phi^{\prime} _A\bzk}}
{{\tilde \delta G[\sigma]} \over {\tilde \delta 
\Pi^{\prime} _A\bzk}}-(-)^{\left| A \right|}{{\tilde \delta F[\sigma]} \over {\tilde \delta \Pi^{\prime} _A\bzk}}
{{\tilde \delta G[\sigma]} \over {\tilde \delta \phi^{\prime} _A\bzk}}} \right).
\end{eqnarray*}
This is easily verified with substituting (\ref{eq:ctransf1}) and (\ref{eq:ctransf2}) into (\ref{eq:pb}).
We can also show that the transformation, which makes the four-dimensional Poisson bracket invariant, 
is exactly the canonical transformation. 

We may express Hamilton's equations (\ref{eq:cano1a}) and (\ref{eq:cano2a}) by the four-dimensional Poisson bracket:
\begin{eqnarray}
  (-)^{\left| A \right|}\partial _n\phi _A \bzk &=& \left[ {\phi _A \bzk,\ H} \right]_c, \label{eq:cano1b}\\
  (-)^{\left| A \right|}\partial _n\Pi _A \bzk  &=& \left[ {\Pi _A \bzk,\ H} \right]_c. \label{eq:cano2b}
\end{eqnarray}

\section {The fundamental algebraic relations}

Let us consider two examples of infinitesimal transformations\par

\noindent
(1) the space-time translation 
\begin{eqnarray*}
\delta x_\mu &=&\varepsilon _\mu ,\\
\delta \phi _A\bxk&=&0,
\end{eqnarray*}
(2) the Lorentz transformation
\begin{eqnarray*}
\delta x_\mu &=&\varepsilon _{\mu \nu }x^\nu ,\\
\delta \phi _A\bxk &=& {i \over 2}\left( {S_{\mu \nu }} \right)_{AB}\phi _B\bxk\varepsilon ^{\mu \nu },
\end{eqnarray*}
where $S_{\mu \nu}$ is the spin matrix.

The generating functions for those transformations are then given by
\begin{eqnarray*}
P_\nu &=& \int_\sigma  {d\Sigma ^\mu }\bxk\ T _{\mu\nu} \lxr,\cr
M_{\mu \nu }&=& \int_\sigma  {d\Sigma ^\lambda }\bxk\ m_{\lambda ,\mu \nu }\lxr,
\end{eqnarray*}
where
\begin{eqnarray*}
M_{\mu \nu }&=& -M_{\nu \mu },\\
m_{\lambda ,\mu \nu }\lxr &=&
x_\mu T_{\lambda \nu }-x_\nu T_{\lambda \mu }+i\ \Pi_{\lambda A}\bxk\left( {S_{\mu \nu }} \right)_{AB}\phi _B\bxk,\\
\Pi_{\lambda A}\bxk &=& {{\partial ^R {\cal L}\lxr} \over {\partial [\partial ^{\lambda}\phi _A\bxk]}}.
\end{eqnarray*}
The ten generating functions $P_{\nu}$ (the total momentum) and
$M_{\mu \nu}$ (the total angular momentum) satisfy the Poincar{\'e}
algebra with the four-dimensional Poisson bracket
\begin{eqnarray*}
\left[ {P_\mu ,P_\nu } \right]_c &=& 0, \\
\left[ {P_\mu ,M_{\lambda \nu }} \right]_c &=&
-g _{\mu \lambda }P_\nu +g _{\mu \nu }P_\lambda ,\\
\left[ {M_{\lambda \kappa },M_{\mu \nu }} \right]_c &=&
-\left( {g _{\nu \lambda }M_{\mu \kappa }+g _{\mu \lambda }M_{\kappa \nu }+g _{\kappa \nu }M_{\lambda \mu }
+g _{\mu \kappa }M_{\nu \lambda }} \right).
\end{eqnarray*}
For the canonical variables on the space-like hypersurface $\sigma$ passing through a point $x$, 
it follows that
\begin{eqnarray}
\left[ {\phi _A\bxk,P_\nu } \right]_c &=& -\partial _\nu \phi _A\bxk,\label{eq:heisenberg1} \\
\left[ {\Pi _A\bxk,P_\nu } \right]_c &=& -\partial _\nu \Pi _A\bxk,\label{eq:heisenberg2} \\
\left[ {\phi _A\bxk,M_{\mu \nu }} \right]_c &=& x_\mu \partial _\nu \phi _A\bxk-x_\nu \partial _\mu \phi _A\bxk
                                                 +i\left( {S_{\mu \nu }} \right)_{AB}\phi _B\bxk,\label{eq:heisenberg3} \\
\left[ {\Pi _A\bxk,M_{\mu \nu }} \right]_c &=& (x_\mu \partial _\nu -x_\nu \partial _\mu )\Pi _A\bxk
                                                -i\Pi _B\bxk\left( {S_{\mu \nu }} \right)_{AB}\nonumber\\
                                           & &+n_\mu \bxk \Pi_{\nu A}-n_\nu \bxk \Pi_{\mu A}. \label{eq:heisenberg4}
\end{eqnarray}

To pass over to quantum field theory in the Heisenberg picture, we shall make all
canonical variables into operators satisfying commutation relations (commutators)
corresponding to the four-dimensional Poisson bracket as
$$
{\hbox{(four-dimensional Poisson bracket)}} \rightarrow {(i\hbar)}^{-1}{\hbox{(four-dimensional commutator)}}.
$$
This is the canonical quantization with covariant manner. The commutation
relations corresponding to (\ref{eq:heisenberg1})--(\ref{eq:heisenberg4}) coincide with the results by Schwinger,
who derived them from the dynamical principle~\cite{SW2}.
%
\section {The symplectic structure}

We go back to the four-dimensional Poisson bracket to handle it in the differential form. 
The four-dimensional Poisson bracket is expressed as
\begin{eqnarray}
\left[ {F[\sigma ],G[\sigma ]} \right]_c 
&=&\!\int_\sigma \!{d\Sigma \left( {{{\tilde \delta F[\sigma ]} \over {\tilde \delta \phi _A}}
	{{\tilde \delta G[\sigma ]} \over {\tilde \delta \Pi _A}}-(-)^{\left| A \right|}
	{{\tilde \delta F[\sigma ]} \over {\tilde \delta \Pi _A}}
	{{\tilde \delta G[\sigma ]} \over {\tilde \delta \phi _A}}} \right)}
\nonumber\\
&=&\!\int_\sigma \! {d\Sigma \! \left( {\bar \delta \phi _A({\bf X}_F)\ \bar \delta \Pi _A({\bf X}_G)
	-(-)^{\left| A \right|}\bar \delta \Pi _A({\bf X}_F)\bar \delta \phi _A({\bf X}_G)\ } \right)}, \nonumber\\
\label{eq:symp1}
\end{eqnarray}
where ${\bf X}_U$ is a vector field
\begin{eqnarray*}
{\bf X}_U &=& \Omega_{mn}{{\tilde \delta U} \over {\tilde \delta z_m}}{{\tilde \delta } \over {\tilde \delta z_n}},\\
\Omega &=& \left( {\matrix{0&1\cr{(-)^{\left| A \right|+1}}&0\cr}} \right),
\end{eqnarray*}
and
$\bar\delta z_{m} ({\bf X}_U)$ is the interior product of $\bar\delta z_m$ and ${\bf X}_U$,
with $z_m = \phi_m$ for $m \leq N$ and $z_m = \Pi_{m-N}$ for $m>N$.
We may abbreviate (\ref{eq:symp1}), with the closed two-form $\tilde\omega = \bar\delta \phi_A \wedge \bar\delta \Pi_A$,  as
\begin{equation}
\left[ {F[\sigma],G[\sigma]} \right]_c=\int_{\sigma} {d\Sigma \ \tilde \omega ({\bf X}_F,{\bf X}_G)}. \label{eq:symp2}
\end{equation}
It is apparent that the four-dimensional Poisson bracket (\ref{eq:symp2}) is invariant under
any canonical transformation because of the property of the closed two-form:
$\tilde\omega = \bar\delta \phi_A \wedge \bar\delta \Pi_A = 
\bar\delta \phi^{\prime}_A \wedge \bar\delta \Pi^{\prime}_A$,
where ${\phi^{\prime}}_A$ and ${\Pi^{\prime}}_A$ represent new canonical variables.

According to the definition of the Lorentz covariant momenta on a space-like hypersurface, 
the right hand side of (\ref{eq:symp1}) can be written as
$$
\int_{\sigma} {d\Sigma^\mu (\bar \delta \phi _A\wedge \bar \delta {\Pi_{\mu A}})}({\bf X}_F,{\bf X}_G).
$$
Thus, we get
\begin{equation}
\left[ {F[\sigma],G[\sigma]} \right]_c=\omega ({\bf X}_F,{\bf X}_G), \label{eq:symp3a}
\end{equation}
where
\begin{equation}
\omega = \int_{\sigma} {d\Sigma ^\mu }\bar \delta \phi _A\wedge \bar \delta {\Pi_{\mu A}}
       = \int_{\sigma} {d\Sigma ^\mu }\bar \delta [\delta x {\cal L}-j_{\mu}]
       =-\int_{\sigma} {d\Sigma^\mu} \bar\delta {\cal J}_{\mu}, \label{eq:symp3b}
\end{equation}
with
$$
{\cal J}_\mu ={{\partial^R {\cal L} } \over {\partial [\partial ^\mu \phi _A ]}}{\bar\delta}
 \phi _A ,
$$
which is called the symplectic current.
The closed two-form (\ref{eq:symp3b}) coincides with the symplectic structure by the geometrical approach~\cite{CrWitt}.

It is well-known that the equivalence of the various methods for defining a
Poisson bracket structure among the observables~\cite{canonical}: namely, the Dirac method~\cite{Dirac}, 
the Pierles method~\cite{Pierles}, 
and the geometrical approach. Our method of using the space-like hypersurface
for defining the four-dimensional Poisson bracket
is also equivalent to those methods through the closed two-form (\ref{eq:symp3b}).
Our four-dimensional Poisson bracket is not at all different in its contents
from other methods, but has the advantage of passing over to quantum
field theory in the Heisenberg picture clearer.

Note that the closed two-form (\ref{eq:symp3b}) is independent of the choice of $\sigma$, but
the four-dimensional Poisson bracket (\ref{eq:symp3a}) is not necessarily so.
This is because $F[\sigma]$ and $G[\sigma]$ may or may not depend on the choice
of $\sigma$. If both $F[\sigma]$ and $G[\sigma]$ are independent of $\sigma$, 
the four-dimensional Poisson bracket will be as well.

%
\appendix\section*{\hbox{   }The method of invariant variation}

Let us consider a set of classical fields $\phi_A \bxk$ and their coordinate derivatives 
$\partial_{\mu} \phi_A\bxk$, with the Lagrangian density 
$${\cal L}(\phi_A \bxk, \partial_{\mu} \phi_A\bxk) \equiv {\cal L}\lxr.$$ 
Now let $I$ be the action integral referred to a space-time domain $\Omega$, 
which is bounded by a piece of space-like hypersurfaces $\sigma_1$ and $\sigma_2$,
$$
I = \int_{\Omega} d^4 x {\cal L}\lxr=
\int_{\sigma _1}^{\sigma _2} {d^4x\ {\cal L}}\lxr. \eqno (A.1)
$$
Then, under the infinitesimal transformation:
$$
x_{\mu} \rightarrow x_{\mu}^{\prime} \equiv x_{\mu} + \delta x_{\mu},
$$
$$
\phi_A\bxk \rightarrow {\phi_A}^{\prime} \bxpk \equiv \phi_A \bxk+ \delta \phi_A\bxk,
$$
the variation of the action integral (A.1) is evaluated as follows
$$
\delta I=\int_\Omega  {d^4}x\ 
\Bigl[\bigl({{\partial^R {\cal L}\lxr} \over {\partial \phi _A \bxk}}
-\partial _\mu {{\partial^R {\cal L}\lxr} \over {\partial [\partial _\mu \phi _A \bxk]}}\bigr)
\bar\delta \phi _A\bxk +\partial _\mu j^\mu \bxk \Bigr], \eqno (A.2)
$$
where
$$
\bar\delta \phi _A \bxk = \delta \phi _A \bxk -  \delta x^\mu \partial_{\mu} \phi_A \bxk,
$$
$$
j^\mu \bxk={{\partial^R {\cal L} \lxr} \over {\partial [\partial _\mu \phi _A \bxk]}}
{\delta} \phi _A \bxk -{T^\mu}_{\nu} \lxr \delta x^\nu, \eqno (A.3)
$$
and
$$
{T^\mu} _\nu \lxr={{\partial^R {\cal L}\lxr} \over {\partial [\partial _\mu \phi _A \bxk]}}
\partial _\nu \phi _A \bxk-{\delta ^\mu }_\nu {\cal L}\lxr.
$$
The $j^{\mu}\bxk$ and  ${T^\mu} _\nu \lxr$ are known as the Noether current and the canonical energy momentum tensor, respectively.

We have adopted the right-differentiation convention $(R)$ with respect to the anticommuting c-number $\xi_A$ 
such as fermion components; that is, the differential operator $(\partial^R / \partial \xi_A)$ has a property
$$
{{\partial ^R(PQ)} \over {\partial \xi_A }}=P\left( {{{\partial Q} \over {\partial \xi_A }}} \right)
+\left( - \right)^{\left| Q \right|}\left( {{{\partial P} \over {\partial \xi_A }}} \right)Q,
$$
where $P$ and $Q$ are any monomials in the commuting and anticommuting c-numbers, 
the symbol $(-)$ means $-1$, and $\left| Q \right|$ is equal to $1$ anticommuting with $\xi_A$ contained in $Q$, else $0$.

If the action integral is invariant, $\delta I=0$,  under the infinitesimal transformation 
and if the field equation holds, equation (A.2) implies the Noether current $j^\mu \bxk$
to be conserved 
$$
\partial_{\mu} j^{\mu} \bxk =0.
$$ 
Then the corresponding conserved charge $J[\sigma]$
$$
J[\sigma]=\int_\sigma  {d\Sigma _\mu \bxk} j^{\mu} \bxk
$$
generates the infinitesimal transformation.
The conservation law $J[\sigma_1]=J[\sigma_2]$ is easily justified with the help of Gauss's theorem
$$
\int_\Omega  {d^4}x\partial _\mu j^\mu \bxk =\int_{\partial \Omega } {d\Sigma _\mu }\bxk j^\mu \bxk,
$$
where $\partial \Omega = \sigma_2 -\sigma_1$ which is made by joining together upper surface $\sigma_2$ 
and lower surface $\sigma_1$.


\begin{center}
\epsfxsize=4cm
\epsfbox{06fig2.eps}
\end{center}
\begin{figure}
\caption{The deformation of the surface $\sigma$ is given by that of the function
$g$ for a fixed parameter $c$. The upper surface $\sigma +\delta\sigma$ is identical with 
$\sigma$ except a small region surrounding a given point.}
\label{fig2}
\end{figure}
	
\end{document}